# A Microwell-Based Microfluidic Device for Single-Cell Trapping and Magnetic Field Gradient Stimulation

Richard Lee Lai[a]

[a] *Graduate Institute of Biomedical Electronics and Bioinformatics, National Taiwan University, Taiwan*

## Abstract

We develop a microfluidic platform for the long-term cultivation and observation of both THP-1 cells under different physiological conditions. First, we determine optimal seeding conditions and microwell geometry. Next, we observe changes in cell size and circularity. Results show that gradient magnetic forces on the order of $10^2$ T/m results in stunted growth and irregular cell shapes. Finally, we observe the temporal change in ROS signals under control, static and gradient magnetic fields. For exposure to static and gradient magnetic fields, the peak in ROS signals occurs after 24 hours and 36 hours, respectively.

## 1. Introduction

Long-term culture and observation of mammalian cells is vital for the understanding of how they behave in actual physiological systems. This so-called *in vitro* or cellular study provides a strong foundation to elucidate pathogenesis of certain condition, and results in possibilities to develop therapeutic interventions. Currently, most studies utilize petri dishes or culture flasks for this purpose[1]. However, these tools require large volumes of medium and are unable to provide the three dimensional structure crucial for the growth of certain cells such as osteocytes and myocytes.[2] Manipulation of individual cells is manually involved and inaccurate, while automation of cell handling is difficult due to the low parallelization of assays. Besides, it is also difficult to isolate single cells for observation, which is important for diagnosis of cancer and immune status[3].

**1.1 Single-Cell Trapping with Microwells**

Due to its precise fluidic control, smaller footprint and high-biocompatibility, microfluidics provide an ideal platform for efficient cell trapping, isolation, and on-chip phenotyping analysis[4]. By using microfluidic devices, it is possible to provide precise control of biophysical conditions, such as shear stress or external stimuli. However, most microfluidic devices focus on the filtering and capture of targeted cells from a sample without providing any means of long-term cultivation and observation.[5]

Currently, there are two major microfluidics-based cell trapping methods: labeled and label-free. For the former, specific fluorescent labeling or immunomagnetic beads were bound to targeted cells for magnetic[6] or optical trapping.[7,8] Label free methods such as optical tweezing, dielectrophoresis and hydrodynamics are based mainly on variances in cell size, density, compressibility and dielectric constants. Of these three, hydrodynamic methods are the most common since it allows for simple device fabrication and sample preparation.

Microfluidic-based hydrodynamic trapping can be mainly classified into lateral and vertical trapping.[9] For lateral trapping, previous studies have demonstrated U-shaped microstructures, C-shaped rings with microsieves, jail bars and a side channel trapping array.[9] For vertical trapping, microwells have been proven to be the most effective and versatile. Compared to lateral trapping, microwells can provide an enclosed microenvironment for on-chip cell culture and stimulation once trapping is achieved.

To increase the number of cells trapped within our device, we aim to improve the well occupancy rate. This can be accomplished using different microwell geometries; previous studies demonstrated that triangular wells have higher trapping rates than circular wells.[10,11] However, we also need to maintain a high initial single cell capture rate to prevent overcrowding of microwells, which may lead to nutrient deficiency in long-term cell culture, lowering cell viability.[12] Therefore, we seek to optimize microwell geometry for both purposes, maximizing the effective cell count.

**1.2 Reactive Oxygen Species (ROS)**

Reactive Oxygen Species (ROS) are chemically reactive compounds containing oxygen. They are formed as a natural byproduct of oxygen metabolism and are important in cell signaling and homeostasis.[13] Cells normally control ROS levels through regulatory pathways and antioxidant molecules such as vitamins A, C, and E.[14] However, under environmental stress such as hypoxia or shear stress, ROS levels may increase dramatically.[15] The resulting DNA damage is responsible for carcinogenesis; cancer cells are found to have elevated levels of ROS close to those that induce programmed cell death.[16] By further increasing ROS levels, it may be possible to induce apoptosis and necroptosis, offering possible treatment. On the other hand, simulating carcinogenic conditions provides insight into the growth and development of cancer cells. Previous studies lacked non-invasive methods to fine-tune ROS levels for this purpose, instead of relying on gene knockouts and pathway manipulation.[17]

Microfluidics has proved to be a versatile tool in the understanding of ROS related phenomena. Lo et.al has reported the use of branched PMMA channels to study lung cancer cells under different chemical and electrical stimuli.[18] Ayuso et.al fabricated a hydrogel-based PDMS device for the detection and monitoring of ROS levels in a tumor microenvironment.[19] Others have designed chips enabling long-term observation of neutrophil extracellular traps (NETs) and the production of ROS in such structures.[20]

**1.3 Classification of Magnetic Fields**

Biological cells respond to stress and strain in their surroundings through mechanosensitivity. To observe the resulting morphological and physiological effects in vitro, researchers have conducted experiments using a variety of optical,[21] hydrodynamic[22], and electrical stimuli.[23] However, these often involve invasive procedures and complicated equipment. Magnetic fields provide label-free and non-invasive stimuli without requiring electrodes or syringe pumps, which are difficult to integrate with microfluidics and increase the overall device footprint. Overall, there are two types of magnetic fields: gradient and alternating. Gradient magnetic fields occur when there is a change in the magnetic field over distance and can be produced by periodic magnets. Previous studies utilized micromagnets fabricated onto silicon wafers to generate magnetic field gradients up to $10^5$ T/m.[24] THP-1 cells were placed in the cavities between magnets and stained with fluorescent ROS dye. Results show increases in ROS and lowered cell growth, suggesting that magnetic field gradients result in greater cellular stress. The intensity of the gradient also has a substantial effect on cell fate. According to previous studies, magnetic field gradients on the order of $10^8$~$10^9$ T/m result in local changes of membrane potential,[25] while those between $10^4$ and $10^6$ T/m can cause cytoskeletal reorganization and assist in cell division and migration.[26] Even smaller gradients such as those around $10^3$ T/m have effects on pathway differentiation[27] and morphological changes such as cell swelling.[24] Such changes are observed over a wide range of cell types such as HeLa cells, THP-1 cells and cancer cells. On the other hand, low-gradient magnetic fields are constant over a given region and are usually generated by commercial NdFeB bulk magnets. They have little effect on cell morphology and proliferation[28] with the exception of cancer cell lines, which exhibit decreased metabolic rate in the case of strong magnetic fields in excess of 1T.

Alternating magnetic fields are the result of either rotating permanent magnets[29] or electromagnetic waves. Santoro et al. reported a decrease in membrane fluidity and reorganization of cytoskeletal components in human lymphoid cells exposed to 50 Hz sinusoidal magnetic field,[30] Here, we focus on using static and gradient magnetic forces since they are more easily generated and do not require EMF shielding during experiments. However, these studies lack the precision required to allow for rapid analysis of cell morphology changes and lack identical physiological flow conditions. Therefore, we combine microfluidics with high-gradient magnetic fields generated by micrometer-scale magnets to address these unmet needs.

**1.4 How Gradient Magnetic Fields Affect Cells**

Transient Receptor Potential (TRP) channels are a family of ion channels that occur mostly on the cellular membrane of animal cells.[31] They are responsible for a variety of sensory functions such as heat detection, gustation and mechanical stress and can be grouped into six subfamilies: TRPC (canonical), TRPV (vanilloid), TRPM (melastatin), TRPA (ankyrin), TRPML (mucolipin), and TRPP (polycystin). Of these, TRPV4[32] has been reported to be activated by both ROS and mechanical forces and are present in THP-1 cells. TRPV4 mediates the influx of extracellular $Ca^{2+}$, causing changes in RNA transcription, vesicular transport and cytoskeleton remodeling. In immune cells, TRPV4 is responsible for the formation of ROS in response to pro-inflammatory stimuli.[33]

Differences in susceptibility between the cell and the surrounding medium result in a magnetic field gradient, which may exert shear and pressure forces on the cell. According to previous studies regarding THP-1 cells,[24] the resulting shear stress is approximately 0.0263 Pa, while the amount of strain induced is around 26.3 Pa. These values are sufficient to cause morphological changes in cells; previous studies stated the effect of shear stress on the cell membrane regarding the activation of mechanosensitive ion channels such as TRPV4. This results in the imbalance of osmotic and hydrostatic pressures, altering the flux of ions transported through the cell membrane. In this paper, we develop a microwell-based microfluidic device to study ROS secretion and morphological changes of THP-1 cells under different magnetic field conditions. By integrating magnetic chips with microfluidics, it is possible to control magnetic field exposure with both temporal and spatial precision.

## 2. MATERIALS AND METHODS

### 2.1 Microfluidic Device Fabrication

The microfluidic device was made of two polydimethylsiloxane (PDMS) layers: (1) the microchannel layer and (2) the circular and triangular microwell layer with 20µm well depth and a period of 100 µm. The microchannel layer contained branches at the inlet and outlets to allow for homogeneous dispersion of cells during seeding. All components were fabricated by the soft lithography process, similar to our previous papers.[6] [6]Briefly, silicon molds were fabricated using photolithography processes. Then, PDMS (Sylgard-184, Dow Corning) prepolymer with a 1:10 weight ratio of PDMS curing agent to base monomer was mixed with 2.5% of silicone oil to facilitate the subsequent release of patterned PDMS layers. We used a spin coater to control the thickness of the PDMS layer to 250±50 µm, ensuring low chip to chip variation of the effective magnetic field. After placing in a 60 °C oven for 2 hours, fully cured PDMS structures were peeled off from the silicon molds, while excessive PDMS was trimmed using a razor blade. The two PDMS layers were aligned and bonded by oxygen plasma (Plasma Cleaner PDC-001, Harrick Plasma) (45 W for 60 s). The bright field images were taken by 10x (UPlanFl, N.A.= 0.30, Olympus) objective, while the fluorescent images were taken using a BW filter (U-FBW, Olympus).

### 2.2 Cell Culture Conditions

THP-1 cells (TIB-202, ATCC) were cultured using complete growth medium (RPMI 1640, Dow Corning) supplemented with 10% (v/v) heat-inactivated fetal bovine serum (FBS) (10082139, Gibco), 2mM L-glutamine and 35µM β-mercaptoethanol. 100µg/ml of penicillin was added to inhibit bacteria growth. The cell culture environment was maintained at 37℃ with 5% $CO_2$ and 100% humidity. THP-1 cells were sub-cultured after reaching densities of $1.5 \times 10^6$ cells/mL to prevent overcrowding, which lowers overall viability. Cell density and viability were measured by an automated cell counter (TC-20 cell counter, Bio-Rad).

### 2.3 Generation of Magnetic Field

To generate a static magnetic field, we used a NdFeB permanent magnet with a magnetization of 4000 Gauss [10 mm (*W*: width) x 10 mm (*L*: length) x 10 mm (*H*: height)] (Uni-magnet). To generate a magnetic field gradient, we designed a chip with periodic grooves made from a 1:1 mixture of PDMS and ferrite powder [10 mm (*W*: width) x 10 mm (*L*: length) x 2.5 mm (*H*: height)]. The smaller the distance between each groove, the greater the magnetic field gradient. In practice, this value is constrained to around 250µm due to limitations of the CNC drill used to fabricate the PMMA mold. Therefore, we use this value in both our simulations and actual fabrication. Since the ferrite powder has a maximum particle diameter of 10µm, much smaller than the grooves, it is not necessary to take into account the magnetic field of each individual particle.

## 2.4 Magnetic Fields Equation and Measurement

We first simulate the magnetic field strength using the commercial finite element method (FEM) software package COMSOL Multiphysics 5.5. The magnetic field strength is calculated using Gauss' law in its differential form:

$$H = -\nabla V_m \quad (1)$$

$$\nabla \cdot B = 0, \ B = \mu_0 \mu_r H \quad (2)$$

Where $V_m$ is the scalar magnetic potential, H is the magnetic field strength, B is the magnetic flux density, $\mu_0$ is the permeability of free space and $\mu_r$ is the permeability coefficient of the material. For a 1:1 mixture of ferrite powder and PDMS, this value is approximately 4000 H/m. We then determine the magnetic field gradient $\nabla B$ by taking the derivative of B along the axis of interest. The magnetic field gradient force can be calculated as:

$$\vec{f_m} = (\chi_c - \chi_m)\frac{(\vec{B}\vec{\nabla})\vec{B}}{\mu_0} \quad (3)$$

Where $x_m$ is the susceptibility of the medium and $x_c$ is the susceptibility of the cell. The difference between these two values $\Delta x$ determines whether cells are attracted to areas with the highest magnetic field gradients ($\Delta x>0$) or are repelled from them ($\Delta x <0$).

## 2.5 The Operational Protocol of the Magnet Integrated Microwell Device

Microfluidic devices were first sterilized using 75% ethanol for one hour. Any remaining ethanol was removed by washing the device with PBS solution. We then load cells at a density of $5 \times 10^5$ cells/ml into the microwell device using a pipette and leave the tips in the inlet and outlet to allow for addition of fresh medium at 24-hour intervals. Finally, we place bulk magnets or magnetic chips beneath the center of microwell devices designated for static and gradient magnetic field treatment, respectively. Chips removed for imaging are not replaced to prevent contamination of incubation equipment. ROS signal intensity was measured using cell-permeant fluorescent dye ($H_2$DCFDA, Invitrogen). Cell viability was estimated using a live-dead staining kit (Invitrogen) containing Calcein-AM for live cell staining and Propium Iodine for dead cell staining.

## 2.6 ROS Signal Processing

To determine the ROS signal intensity of trapped cells, we need to subtract the grayscale intensity of the cell from that of its surroundings. This is determined by first placing a threshold mask on the entire image, defining each cell as an independent sub-region. The mask is generated using the Canny edge detection method and setting the lower and upper threshold values to 0.04 and 0.4, respectively. We then find the average grayscale intensity of the pixels within each sub-region. Obtaining the intensity of the pixels surrounding each cell is somewhat more difficult; in theory it is necessary to calculate the average intensity of pixels forming a ring around its edge. However, this requires additional detection algorithms to determine whether the ring is entirely within the microwell or overlaps with another cell, which have different background intensity from other regions. To simplify the process, we instead sample pixels along two perpendicular axes, using the centroid of each cell as the origin. Calculation time is further reduced by sampling pixels six to eight pixels away from the edge of each cell, which is enough to mitigate the effect of blurring at the edges. If the cell happens to be close to the microwell edge, there will be a drop in the sampled value. The program will pause and remove this value before switching to another direction. Since most cells are trapped in corners, it is unlikely that all four sampling directions coincide with a microwell edge. A similar method is employed to resolve the situation in which two cells are closely adjacent. Only when this fails to provide reasonable results, such as when an arbitrary cell is entirely surrounded by other cells, will the ring method and the ensuing detection algorithms be invoked.

**2.7 Cell Size and Circularity Calculation**

Determining the effects of magnetic fields on the size and circularity of THP-1 cells is slightly more difficult since this requires using bright-field images. A major problem arises when two cells intersect with one another, causing spatial distortions. This can be solved by removing connected components larger than the average size of a cell. Only when two cells are completely separated will they be taken into consideration. We first use Canny edge detection to determine the edges of an arbitrary microwell and use image dilation to form a mask (Figure S2B). Next, a periodic array of such masks is generated to cover the microwell array and conduct binary subtraction on the original image. This allows the removal of all microwell edges, which may interfere with image processing. Finally, we use the relationship $\frac{4\pi A}{P^2}$ to estimate the circularity of each cell. Here, A is the area of the connected component and P is its perimeter. The calculated value varies between 0 for straight lines and 1 for perfect circles. Connected components with a circularity lower than 0.5 are removed. Normal THP-1 cells tend to be more circular while those under environmental stress may become irregularly shaped due to degradation of the cytoskeleton.

# 3. RESULTS AND DISCUSSION

## 3.1 Magnetic Field and Gradient Simulation

Since the period of the microwells (100 µm) differs from the period of the magnetic field (250 µm), we need to calculate the average magnetic field strength and gradient experienced by each well. According to simulation results, the values are 0.125T and 50 T/m, respectively, with the variation within each microwell less than 25%. To validate these values, we measured the magnetic field using a Linear Hall sensor (Texas Instruments). The sensor was placed on a motorized stage and advanced at a constant velocity along the x-axis of the magnet's surface (Figure S1). An Arduino Uno board was used for sensor calibration and data acquisition. In general, the measurements agreed well with the simulations, with less than 5% variation between measured and simulated values, although the edge effect was not observed in the actual bulk magnet due to its rounded edges. The simulated and measured peak magnetic field strengths for bulk magnets were 0.19T and 0.185T, while for high gradient magnets they were 0.175T and 0.172T, respectively. The smaller values for high gradient magnets were due to the weaker magnetic field of the ferromagnetic layer. Next, we calculate the magnetic field gradients by differentiating along the magnetic field strength curve. The results for bulk and high gradient magnets were 5T/m and 150T/m for the regions of interest, respectively (Figure 1).

## 3.2 Cell Trapping Results Based on Different Geometry and Volume

We conduct a series of experiments to determine the optimal microwell device and cell seeding volume. To simplify image processing, it is expedient to maximize the single cell trapping rate. Multiple occupancy within wells increases the probability that a given cell is confined to corner regions, which reduces the effectiveness of Canny edge detection. At the same time, we also need to increase the total throughput of the device, since this reduces statistical error and is more representative of the entire cell population. Therefore, we define the product of the single cell trapping rate and well occupancy as the single cell occupancy rate and employ it as a figure of merit.

Results show that for all three types of microwells, the greater the cell seeding volume, the higher the single cell occupancy rate (Figure 2). However, for triangular wells with high aspect ratios (W/L=2), a slight decrease in single cell occupancy is registered for seeding volumes greater than 150µL. This is due to the larger cross-sectional area of the microwell in the y-z plane, increasing the probability that multiple cells become trapped in each well. Next, we discuss the relationship between microwell geometry and single cell occupancy. Triangular wells are shown to perform better than circular wells, with the most marked difference for seeding volumes of 150 µL. Therefore, we choose triangular wells with W/L=2 and a seeding volume of 150 µL for our experiments.

## 3.3 Effect of Magnetic Fields on Cell Size and Circularity

We first incubate cells in microfluidic devices for 12, 18, 24 and 36 hours at 37℃ with 5% $CO_2$ and 100% humidity under control, bulk magnet and high gradient magnet conditions, respectively. Results show that for both the control and bulk magnet conditions, cell size increased with incubation period with slightly less growth under static conditions (Figure 3). Under high gradient conditions, cell size was consistently smaller than for other conditions, with the difference between control and high gradient groups rising from 5% at 12 hours to over 25% at 36 hours. This shows that magnetic field gradients inhibit cell growth. Next, we analyze the results for cell circularity. A decrease in circularity suggests differentiation of monocytes into macrophages,[34] which is indicative of increased environmental stress such as inflammation and infection.[35] After 24 hours, there is a significant decrease in the circularity of cells under all three conditions, with those exposed to high gradient magnetic forces experiencing the greatest decline. At 36 hours, there is a slight uptick in the circularity of cells in both the control and bulk magnet group, but not for those in the high gradient group. The observed recovery can be attributed to cell proliferation, since THP-1 cells undergo mitosis about every 26 hours.[36] This reduces the percentage of cells with low circularity and results in larger error bars for longer incubation time periods. For high gradient magnetic fields, lowered cell proliferation may prevent such replacement.

## 3.4 Effect of Magnetic Fields on ROS

Finally, we determine the effect of different incubation periods on the ROS signal intensity. For both bulk magnet and high gradient magnetic field conditions, the ROS signal increases with incubation time, while for the control group it is nearly constant (Figure 4). At 36 hours, the ROS signal for cells exposed to high gradient magnetic fields is 3.5 times that of the control group, while for bulk magnets it is only 30 percent higher. Notice that the greatest increase in the ROS signal occurs between 18 to 24 hours for gradient fields but occurs between 24 and 36 hours for the static one. Since the activation of ROS pathways is the greatest 12~24 hours after exposure to external stress,[37] it is possible that static magnetic fields result in delayed activation of the relevant pathways.

To elucidate the effects of longer magnetic field exposure, we also incubated the devices for 48 hours. Results show that the ROS signal under gradient conditions declined to levels similar to those under normal conditions, while those under static conditions continued to rise. Previous studies had incubation times of less than 24 hours, which may explain the absence of such observations.[24,25] This not only confirms the delayed activation hypothesis mentioned in the results section, but also suggests the evidence of mechanosensory fatigue in TRPV4 channels, which reduces the sensitivity of cells to environmental changes over time.

**3.5 Discussion**

At the same time, we also conducted cell viability tests, but found no statistically significant results. In fact, it appears that the viability for cells exposed to gradient magnet conditions is slightly higher than those in the control group. It seems that a magnetic field gradient on the scale of $10^2$ T/m is sufficient to cause changes in cellular ROS and morphology, while having minimal effect on cell viability. Another interesting thing to note is that instead of cell swelling as observed previously,[24] our results showed decreases in cell size for bulk and high gradient magnet conditions relative to the control group. This is probably due to the differences in the strength and gradient of the magnetic fields used; smaller magnetic field gradients may be sufficient to slow cell growth without causing acute injury, the major cause of swelling.

# 4. CONCLUSION

Overall, our device demonstrates a robust platform for precise spatial and temporal high-throughput single-cell analysis without the need for bulky equipment and large sample volumes. We showed that the application of a high gradient magnetic field induces significant changes in cell morphology and ROS secretion. In the future, we aim to observe the effect of different magnetic field gradient intensities and optimize the exposure time for specific pathways. Another possible avenue of research is further multiplexing of our device; by designing additional side chambers, it is also possible to observe cell proliferation and reduce disruption during the loading of medium or fluorescent dyes. The image detection algorithm can also be further improved by using deep learning methods to determine cell viability through morphological changes. Since Calcein-AM has the same fluorescent emission spectrum as $H_2DCFDA$, this allows us to simultaneously observe ROS signal intensity and viability.

**Acknowledgments:** The author will like to thank Mr. Zheng-Jie Liao and Mr. Ping-Fan Chen regarding wafer fabrication. He will also like to thank Mr. Nien-Tsu Huang for his support regarding funding. This work was financially supported by the "Center for electronics technology integration (NTU-108L900502)'' from The Featured Areas Research Center Program within the framework of the Higher Education Sprout Project by the Ministry of Education (MOE) in Taiwan.

**Data Availability Statement:** The data that supports the findings of this study are available within the article and its supplementary material.

**Author Contributions:** In this work, Mr. R. Lai designed the experiments, fabricated the microfluidic device, performed the experiment, data analysis, and wrote the manuscript.

**Conflicts of Interest:** The author declares no conflict of interest. Also, the founding sponsors had no role in the design of the study, the collection, analyses or interpretation of data, nor in the writing of the manuscript and in the decision to publish the results.

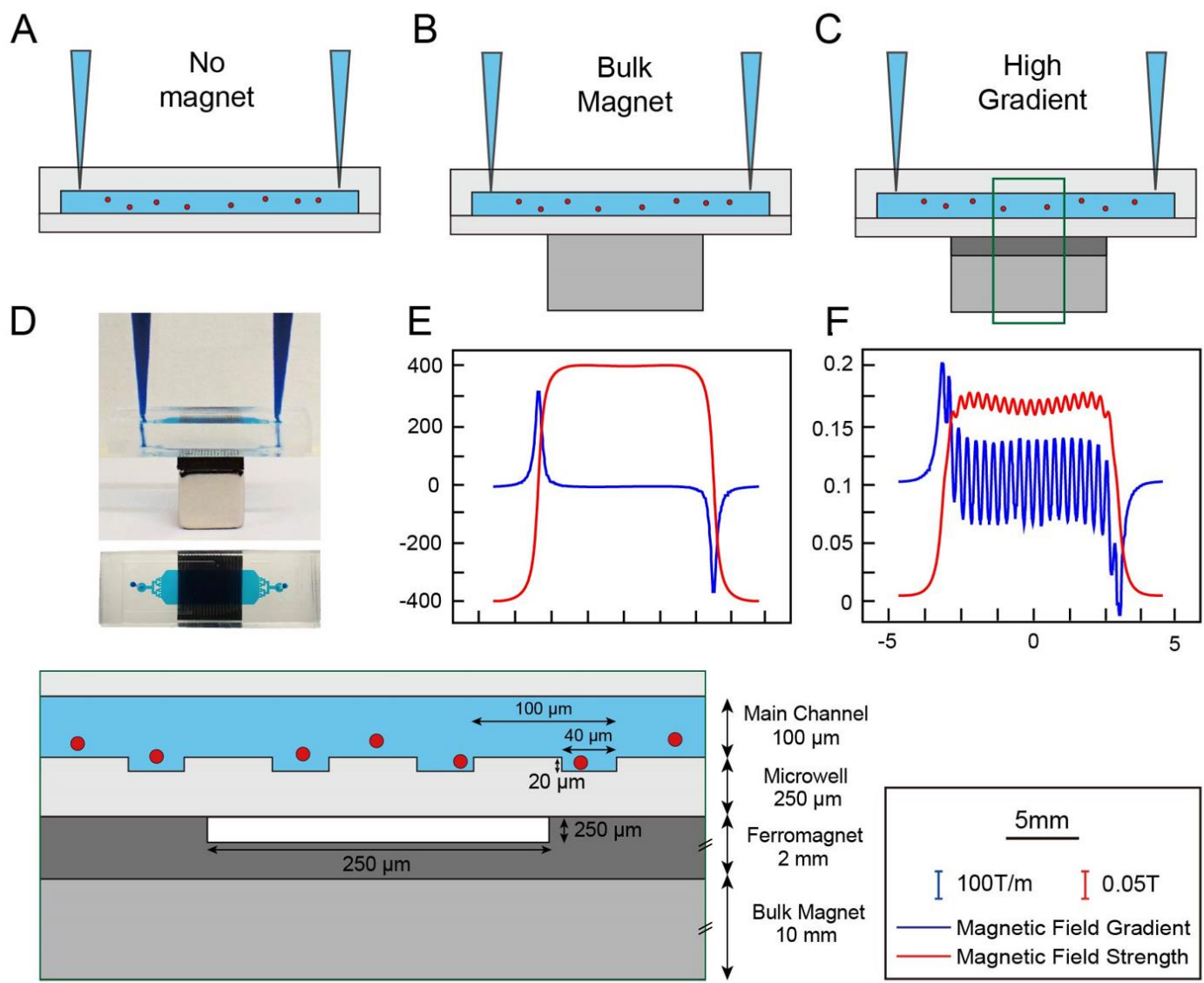

Figure 1. System layout and simulation results. The inset at the right shows the actual device dimensions. The period of the microwell array is 100μm, while that of the ferrimagnet is 250μm. Each microwell has a length of 40μm and a depth of 20μm. The microwell layer thickness is controlled by spin-coating to an average of 250μm. Schematic of experimental setup for device under (A) no magnet (B) bulk magnet and (C) high gradient magnetic field conditions. (D) Actual layout of microfluidic device, which consists of a main channel (colored using food dye) and a microwell layer. Pipette tips in the inlet and outlet allow simple loading of additional nutrients and fluorescent dyes. For the high gradient case, the magnetic layer consists of the ferromagnetic chip and a bulk magnet. (E)(F) The blue and red lines represent the simulated magnetic field gradient and magnet field strength, respectively.

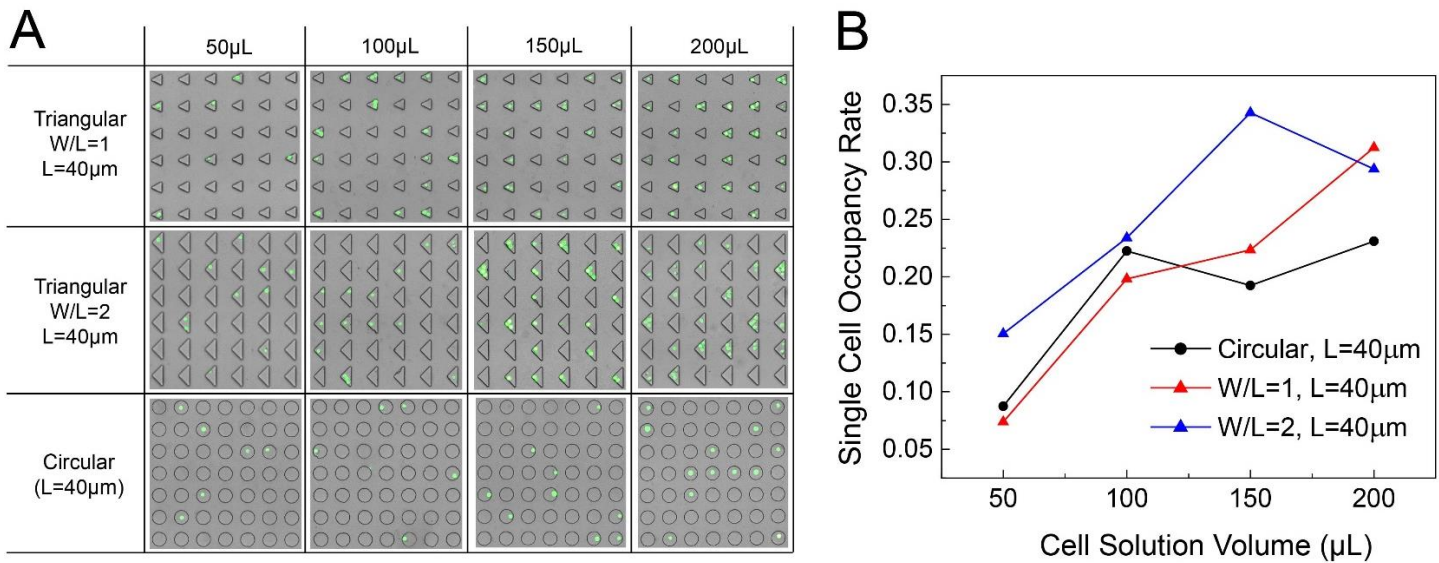

Figure 2. Effect of microwell geometry and solution volume on cell trapping efficiency. (A) Calcein-AM stained fluorescent images of cells trapped within microwells. (B) Triangular microwells with W/L=2 have higher trapping efficiency than those with W/L=1. Circular wells perform more poorly than both kinds of triangular wells. Increasing the cell solution volume results in greater single cell occupancy rate for both circular and W/L=1 triangles. For W/L=2, the maximum rate occurs when the cell solution volume is 150 µL.

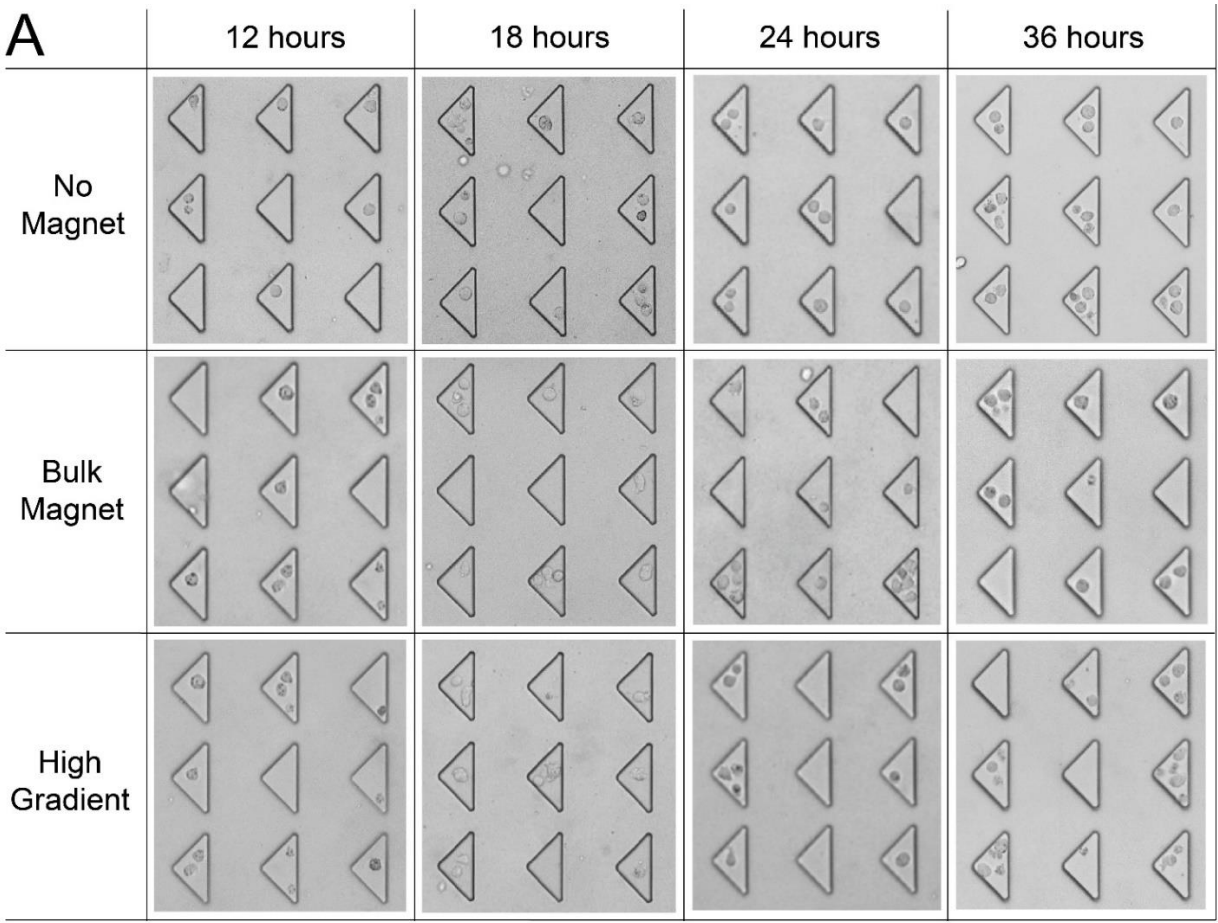

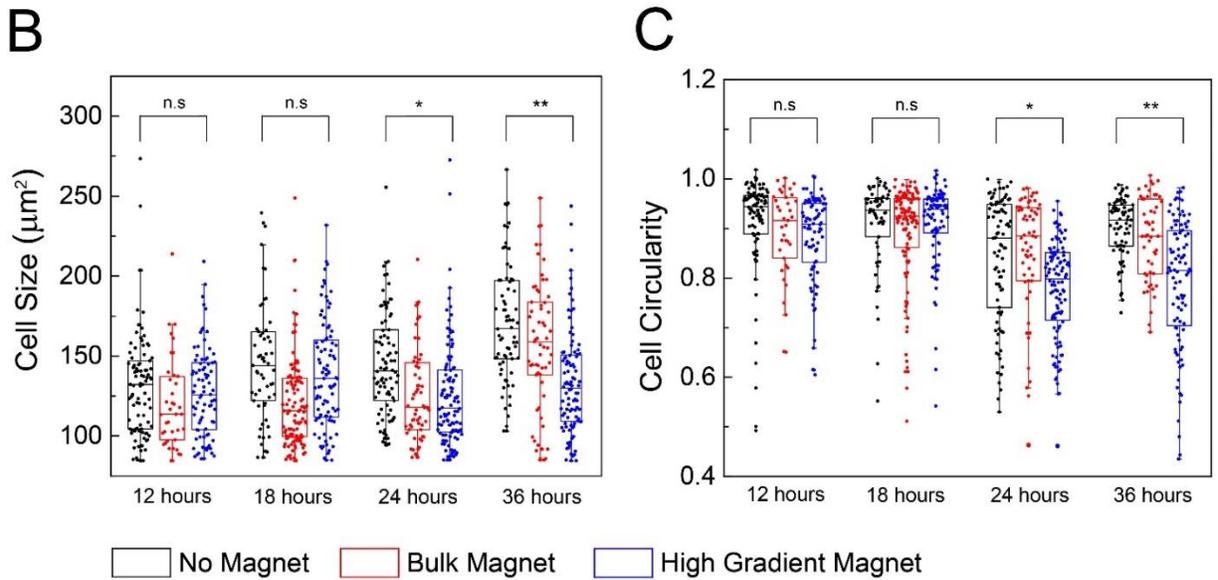

Figure 3. (A) Bright field image of microwells and trapped cells. (B)(C) Effect of magnetic fields on cell size and circularity, respectively. For both the control and bulk magnet groups, cell size increased with incubation time, while for high gradient magnetic fields there was little difference. There is a marked decrease in cell circularity under gradient conditions after 24 hours. For all figures, n=50~100 cells. *: $p<0.01$, **: $p<0.001$, n.s: not significant.

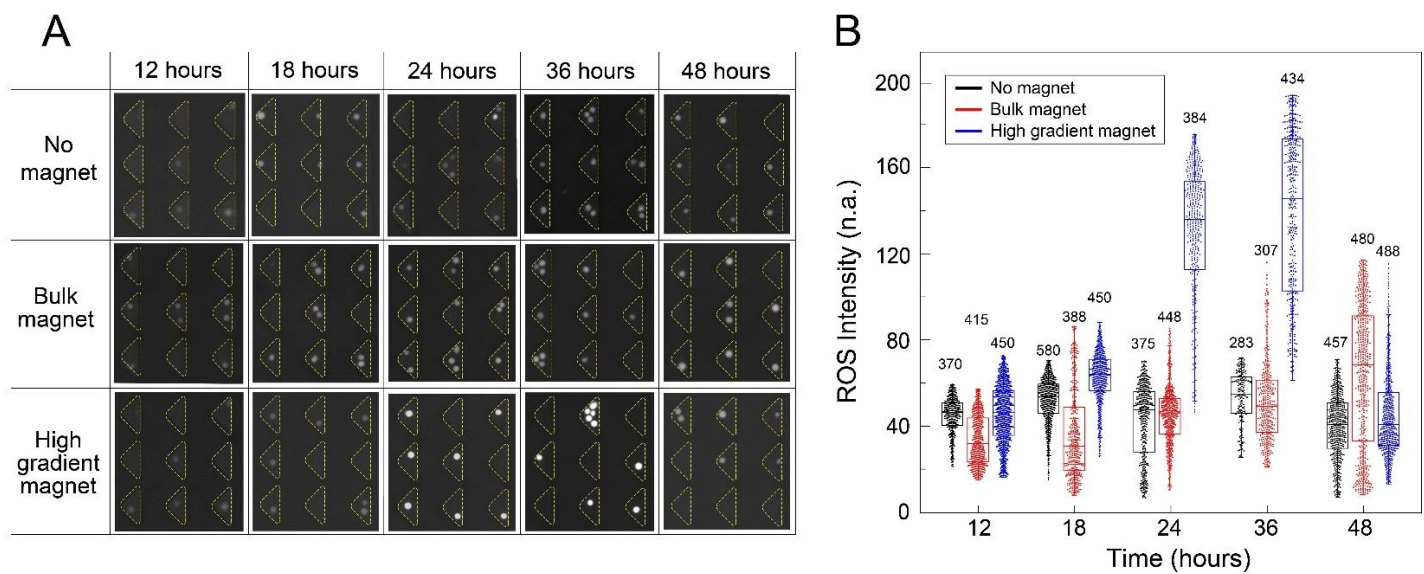

Figure 4. Effect of magnetic fields on ROS signal contrast. (A) Fluorescent image of ROS-stained There is a marked increase in the ROS signal strength under gradient conditions after 24 hours, and little or no change for the control group. For all figures, n=500~1500 cells and p<0.001.

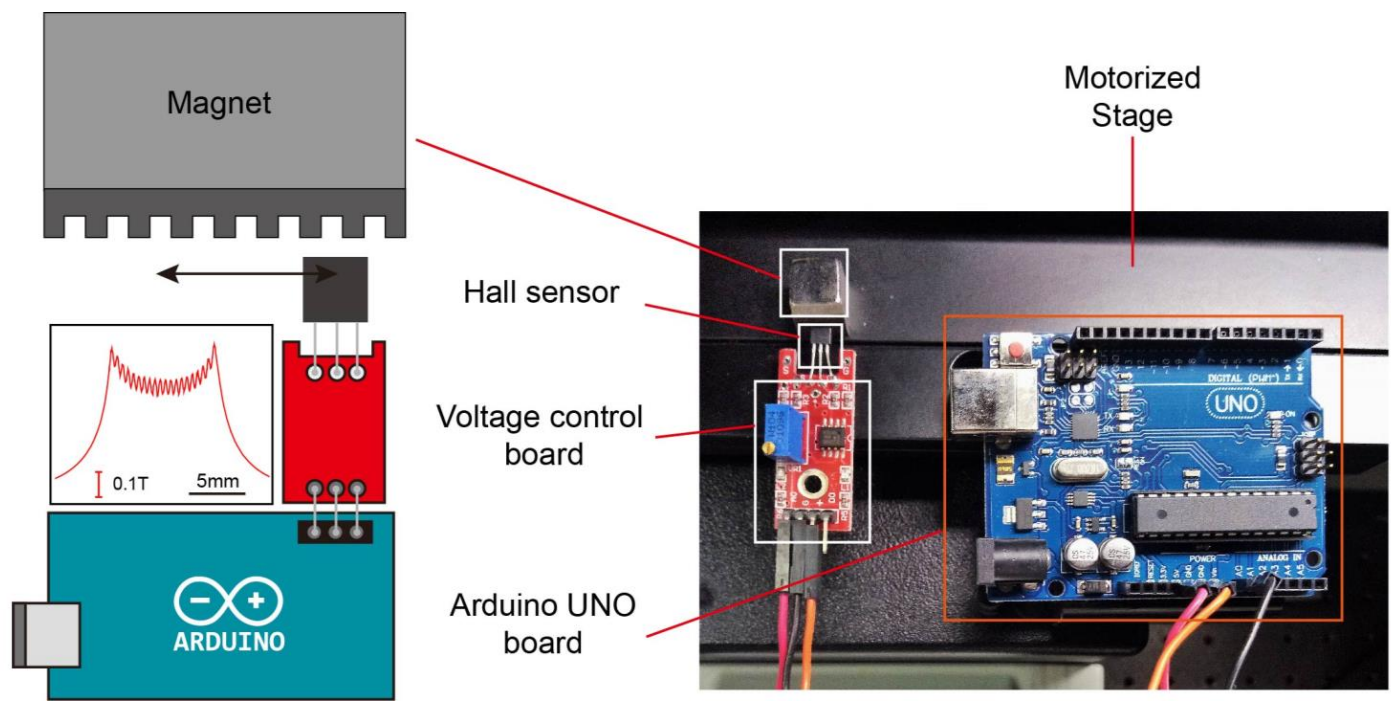

Figure S1. Schematic of apparatus used to measure magnetic field strength. We measured the magnetic field using a Linear Hall sensor. The sensor was placed on a motorized stage and advanced at a constant velocity along the x-axis of the magnet's surface. An Arduino Uno board was used for sensor calibration and data acquisition.

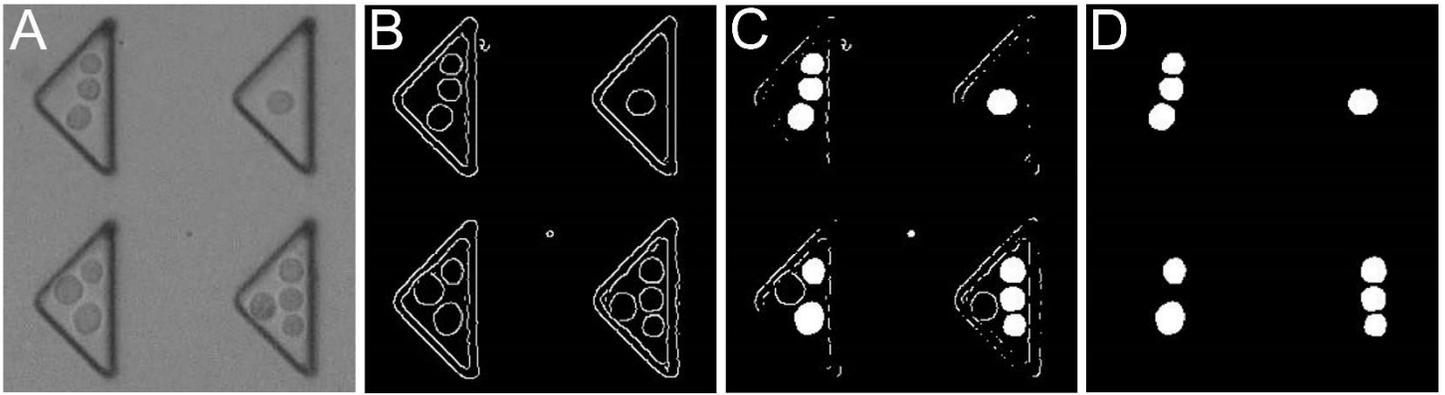

Figure S2. Image processing method used in our study. (A) Bright field image (B) Results of Canny edge detection (C) Cells detected in the figure (D) Final image.